\documentclass[preprint,showpacs,preprintnumbers,amsmath,amssymb]{revtex4}
\usepackage{booktabs}
\usepackage{mathrsfs}
\usepackage{epsfig}
\usepackage{graphicx}% Include figure files
\usepackage{dcolumn}% Align table columns on decimal point
\usepackage{bm}% bold math
\usepackage{amsmath}
\usepackage{slashed}
\usepackage{multirow}

\let\jnfont=\rm
\def\NPB#1,{{\jnfont Nucl.\ Phys.\ B }{\bf #1},}
\def\PLB#1,{{\jnfont Phys.\ Lett.\ B }{\bf #1},}
\def\EPJC#1,{{\jnfont Eur.\ Phys.\ Jour.\ C }{\bf #1},}
\def\PRD#1,{{\jnfont Phys.\ Rev.\ D }{\bf #1},}
\def\PRL#1,{{\jnfont Phys.\ Rev.\ Lett.\ }{\bf #1},}
\def\MPLA#1,{{\jnfont Mod.\ Phys.\ Lett.\ A }{\bf #1},}
\def\JPG#1,{{\jnfont J.\ Phys.\ G}{\bf #1},}
\def\CTP#1,{{\jnfont Commun.\ Theor.\ Phys.\ }{\bf #1},}
\def\ZPC#1,{{\jnfont Z.\ Phys.\ C }{\bf #1},}
\def\JHEP#1,{{\jnfont JHEP \ }{\bf #1},}
\def\Rv{\not{\hbox{\kern-1pt $R$}}}
\def\p{\not{\hbox{\kern-3pt $p$}}}

\newcommand{\met}{\not\!\!\!E_{T}}

\begin{document}

\title{Search for $thj$ production with $h\rightarrow \gamma\gamma$ at the LHC in the
littlest Higgs model with T-parity}% Force line breaks with \\
%%\thanks{A footnote to the article title}%

\author{Bingfang Yang}\email{yangbingfang@htu.edu.cn}
\author{Biaofeng Hou}
\author{Huaying Zhang}
\affiliation{$^1$ College of Physics and Materials Science, Henan
Normal University, Xinxiang 453007, China
   \vspace*{1.5cm} }%

\date{\today}% It is always \today, today,

\begin{abstract}

In the littlest Higgs model with T-parity, we study associated
production of a Higgs and a single top quark at the 14 TeV LHC. We
focus on the Higgs to two photons decay and the semileptonic top
decay channel. By performing a fast detector simulation, we find
that the $thj$ search in the selected channel can excluded the top
partner mass $m_{T_{+}}$ up to 805 (857) GeV for case A (case B) at
$2\sigma$ confidence level at 14 TeV LHC with the integrated
luminosity $L= 3 \rm ab^{-1}$.

\end{abstract}
 \maketitle

\section{INTRODUCTION}

The discovery of a Higgs boson with the CMS and ATLAS experiments in
2012 \cite{higgs-lhc} opened a new field for explorations in the
realm of particle physics. In order to test whether the Higgs boson
is the one predicted by the Standard Model (SM), it is critical to
explore the coupling of this Higgs boson with the other elementary
particles. In particular, the Yukawa coupling of the Higgs to
fermions is an important class, where the top quark owns the
strongest Yukawa coupling due to the large mass. So, it is widely
believed that the top quark is sensitive to the electroweak symmetry
breaking mechanism and new physics\cite{topphy}.

As a direct probe of the top Yukawa coupling, the Higgs boson
production in association with top-quark pair is a golden channel.
But, it is only sensitive to the magnitude of the top Yukawa
coupling rather than its sign. As a beneficial supplement, the
production of a Higgs boson in association with single top quark can
bring a unique possibility\cite{thj-theory}. Higgs boson plus single
top quark production proceeds through three modes, that is
$t$-channel($pp\rightarrow thj$), $s$-channel($pp\rightarrow
th\bar{b}$) and $W$-associated production ($pp\rightarrow thW$),
where the $t$-channel is the dominant mode. Recently, this process
has been measured by the CMS experiments\cite{th-exp}. Meanwhile,
the relevant phenomenological studies have been carried out
extensively\cite{th-theory}.

The littlest Higgs model with T-parity (LHT)\cite{LHT} was proposed
as a possible solution to the hierarchy problem and so far remains a
popular candidate of new physics. The LHT model predicts new gauge
bosons, scalars and top partner, they are responsible for canceling
the quadratic divergence contribution to Higgs boson mass from the
SM gauge boson loops, Higgs self-energy and top quark loop,
respectively. These new particles may contribute to the
$pp\rightarrow thj$ process. Besides, the Higgs couplings are
modified with respect to their SM values and this effect can also
influence the process $pp\rightarrow thj$. By performing the
detailed analysis on the process $pp\rightarrow thj$ may provide a
good opportunity to probe the LHT signal, and the Higgs to
$b\bar{b}$ decay channel has been studied in our previous
work\cite{thjLHT}. In this paper, we will focus on this process and
investigate the observability of $pp\rightarrow thj$ with the
semileptonic decay of the top quark and the diphoton decay of the
Higgs boson. Though the diphoton branching fraction of the Higgs
boson is very small, the diphoton final state allows very good
background rejection thanks to the excellent diphoton invariant mass
resolution provided by the CMS detector.

The paper is organized as follows. In Sec.II we give a brief review
of the LHT model related to our work. In Sec.III we explore the observability of the
process $pp\rightarrow thj$ with $t\rightarrow \ell^{+}\nu b$ and
$h\rightarrow \gamma\gamma$ at 14 TeV LHC
by performing a fast detector simulation and make some discussions. Finally, we give a summary
in Sec.IV.

\section{A brief review of the LHT model}

The LHT model was based on a non-linear $\sigma$ model describing an
$SU(5)/SO(5)$ symmetry breaking, with the global group $SU(5)$ being
spontaneously broken into $SO(5)$ by a $5\times5$ symmetric tensor
at the scale $f\sim \mathcal O$(TeV).

In the fermion sector, the implementation of T-parity requires the
existence of mirror partners for each original fermion. In order to
do this, two fermion $SU(2)$ doublets $q_1$ and $q_2$ are introduced
and $T$-parity interchanges these two doublets. A $T$-even
combination of these doublets is taken as the SM fermion doublet and
the $T$-odd combination is its $T$-parity partner. The doublets
$q_1$ and $q_2$ are embedded into incomplete $SU(5)$ multiplets
$\Psi_1$ and $\Psi_2$ as $\Psi_1 = (q_1, 0, 0_{1\times2})^T$ and
$\Psi_2 = (0_{1\times2}, 0, q_2)^T$. To give the additional fermions
masses, an $SO(5)$ multiplet $\Psi_c$ is also introduced as
$\Psi_c=(q_c,\chi_c,\tilde{q}_c)^T$, where $\chi_c$ is a singlet and
$q_c$ is a doublet under $SU(2)$. Their transformation under the
$SU(5)$ is non-linear: $\Psi_c \to U\Psi_c$, where $U$ is the
unbroken $SO(5)$ rotation in a non-linear representation of the
$SU(5)$. The components of the latter $\Psi_c$ multiplet are the
so-called mirror fermions. Then, one can write down the following
Yukawa-type interaction to give masses of the mirror fermions
\begin{eqnarray}
\mathcal{L}_{\textrm{mirror}}=-\kappa_{ij}f\left(\bar\Psi_2^i\xi +
  \bar\Psi_1^i\Sigma_0\Omega\xi^\dagger\Omega\right)\Psi_c^j+h.c.\
\end{eqnarray}
where $\Omega= \rm diag(1, 1, -1, 1, 1)$, $i, j=1,2,3$ are the
generation indices. The masses of the mirror quarks $u_{H}^{i},
d_{H}^{i}$ and mirror leptons $l_{H}^{i}, \nu_{H}^{i}$ up to
$\mathcal O(v^{2}/f^{2})$ are given by
\begin{eqnarray}
m_{d_{H}^{i}}&=&\sqrt{2}\kappa_if, ~~m_{u_{H}^{i}}=
m_{d_{H}^{i}}(1-\frac{v^2}{8f^2}),\\
m_{l_{H}^{i}}&=&\sqrt{2}\kappa_if, ~~m_{\nu_{H}^{i}}=
m_{l_{H}^{i}}(1-\frac{v^2}{8f^2}),
\end{eqnarray}
where $\kappa_i$ are the diagonalized Yukawa couplings,
$v=v_{SM}(1+\frac{1}{12}\frac{v_{SM}^2}{f^2})$ and $v_{SM}= 246$ GeV
is the vacuum expectation value of the SM Higgs field.

In the top quark sector, two singlet fields $T_{L_1}$ and $T_{L_2}$
(and their right-handed counterparts) are introduced to cancel the
large radiative correction to the Higgs mass induced by the top
quark. Both fields are embedded together with the $q_1$ and $q_2$
doublets into the $SU(5)$ multiplets: $\Psi_{1,t} = (q_1, T_{L_1},
0_2)^T$ and $\Psi_{2,t} = (0_2, T_{L_2}, q_2)^T$. The $T$-even
combination of $q_i$ is the SM fermion doublet and the other $T$-odd
combination is its $T$-parity partner. Then, the $T$-parity
invariant Yukawa Lagrangian for the top sector can be written down
as follow:
\begin{eqnarray}
{\cal L}_t &=& - \frac{\lambda_1 f}{2\sqrt{2}}\epsilon_{ijk}\epsilon_{xy} \left[(\bar{\Psi}_{1,t})_i \Sigma_{jx} \Sigma_{ky} - (\bar{\Psi}_{2,t} \Sigma_0)_i \Sigma^{'}_{jx} \Sigma^{'}_{ky} \right] t^{'}_R \nonumber \\
&& -\lambda_2 f (\bar{T}_{L_1} T_{R_1} + \bar{T}_{L_2} T_{R_2}) +
~{\rm h.c.}
\end{eqnarray}
where $\epsilon_{ijk}$ and $\epsilon_{xy}$ are the antisymmetric
tensors with $i,j,k=1,2, 3$ and $x, y=4, 5$,
$\Sigma^{'}=\langle\Sigma\rangle\Omega\Sigma^\dagger\Omega\langle\Sigma\rangle$
is the image of $\Sigma$ under $T$-parity, $\lambda_1$ and
$\lambda_2$ are two dimensionless top quark Yukawa couplings,
$t^{'}_R$ and $T_{R_{m}}(m=1,2)$ are $SU(2)$ singlets. Under
$T$-parity, these fields transform as: $T_{L_1} \leftrightarrow -
T_{L_2}$, $T_{R_1} \leftrightarrow -T_{R_2}$, $t^{'}_R \to t^{'}_R$.
So, the $T$-parity eigenstates can be defined as
\begin{eqnarray}
t_{L_+}=(t_{L_1}-t_{R_1})/\sqrt{2}, ~~T^{'}_{L_\pm}=(T_{L_1} \mp
T_{L_2})/\sqrt{2}, ~~T^{'}_{R_\pm}=(T_{R_1} \mp T_{R_2})/\sqrt{2}.
\end{eqnarray}
At the tree level, the $T^{'}_{L_-}$ and $T^{'}_{R_-}$ do not mix
with the mirror fermions and the T-odd Dirac fermion
$T_{-}=(T^{'}_{L_-}, T^{'}_{R_-})$ does not interact with Higgs
boson. However, the two T-even eigenstates $(t_{L_+},t^{'}_R)$ and
$(T^{'}_{L_+},T^{'}_{R_+})$ mix with each other so that the mass
eigenstates can be defined as
\begin{eqnarray}
t_L &=& \cos\beta \,t_{L_+} - \sin\beta \,T^{'}_{L_+}, \quad T_{L_+} = \sin\beta \,t_{L_+} +\cos\beta \,T^{'}_{L_+},\nonumber \\
t_R &=& \cos\alpha \,t^{'}_R - \sin\alpha \,T^{'}_{R_+}, ~\quad
T_{R_+} = \sin\alpha \,t^{'}_R + \cos\alpha \,T^{'}_{R_+},\
\label{combination}
\end{eqnarray}
where the mixing angles $\alpha$ and $\beta$ can be defined by the
dimensionless ratio $R=\lambda_1/\lambda_2$ as,
\begin{eqnarray}
\sin\alpha=\frac{R}{\sqrt{1+R^2}}, \quad
\sin\beta=\frac{R^2}{1+R^2}\frac{v}{f}.
\end{eqnarray}
The $t\equiv (t_{L}, t_{R})$ quark is identified with the SM top
quark, and $T_+ \equiv (T_{L_+}, T_{R_+})$ is its T-even heavy
partner, which is responsible for the cancelation of the quadratic
divergence to the Higgs mass induced by the top quark. The heavy
quark $T_{+}$ mix with the SM top quark and leads to a modification
of the top quark couplings with respect to the SM. Then, the masses
of the top quark and its partners up to $\mathcal O(v^{2}/f^{2})$
are given by
\begin{eqnarray}
&&m_t=\frac{\lambda_2 v R}{\sqrt{1+R^2}} \left[ 1 + \frac{v^2}{f^2}
\left( -\frac{1}{3} + \frac{1}{2} \frac{R^2}{(1+R^2)^2} \right)\right]\nonumber \\
&&m_{T_{+}}=\frac{f}{v}\frac{m_{t}(1+R^2)}{R}\left[1+\frac{v^{2}}{f^{2}}\left(\frac{1}{3}-\frac{R^2}{(1+R^2)^2}\right)\right] \nonumber \\
&&m_{T_{-}}=\frac{f}{v}\frac{m_{t}\sqrt{1+R^2}}{R}\left[1+\frac{v^{2}}{f^{2}}\left(\frac{1}{3}-\frac{1}{2}\frac{R^2}{(1+R^2)^2}\right)\right]\label{Tmass}
\end{eqnarray}

The T-invariant Lagrangians of the Yukawa interactions of the
down-type quarks and charged leptons can be constructed by two
possible ways, which are denoted as case A and case B,
respectively\cite{caseAB}. In the two cases, the corrections to the
Higgs couplings with the down-type quarks and charged leptons with
respect to their SM values are given at order $\mathcal{O} \left(
v_{SM}^4/f^4 \right)$ by ($d \equiv d,s,b,\ell^{\pm}_i$)
\begin{eqnarray}
    \frac{g_{h \bar{d} d}}{g_{h \bar{d} d}^{SM}} &=& 1-
        \frac{1}{4} \frac{v_{SM}^{2}}{f^{2}} + \frac{7}{32}
        \frac{v_{SM}^{4}}{f^{4}} \qquad \text{case A} \nonumber \\
    \frac{g_{h \bar{d} d}}{g_{h \bar{d} d}^{SM}} &=& 1-
        \frac{5}{4} \frac{v_{SM}^{2}}{f^{2}} - \frac{17}{32}
        \frac{v_{SM}^{4}}{f^{4}} \qquad \text{case B}
    \label{dcoupling}
\end{eqnarray}

\section{Numerical results and discussions}
\begin{figure}[htbp]
\scalebox{0.55}{\epsfig{file=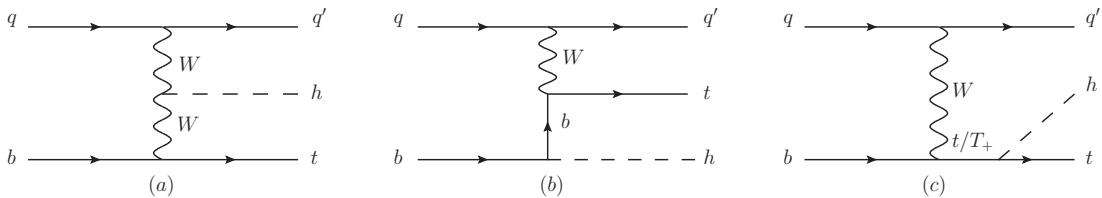}}\vspace{-0.5cm}\caption{
Feynman diagrams for $pp \to thj$ in the LHT model at tree
level.}\label{thjlht}
\end{figure}

In the LHT model, the tree-level Feynman diagrams of the process $pp
\to thj$ are shown in Fig.\ref{thjlht}, where the Higgs is emitted
mainly either from a top quark leg or a $W$ boson propagator. Due to
the couplings of the Higgs to the $W$ and the top quark have
opposite sign, these two diagrams suffer from destructive
interference. The T-even top partner $T_{+}$ contributes this
process through Fig.\ref{thjlht}(c), which will reflect the LHT
effect.

%%Fig.1 %%%%%%%%%%%%%%%%%%%%

\begin{figure}[htbp]
\scalebox{0.35}{\epsfig{file=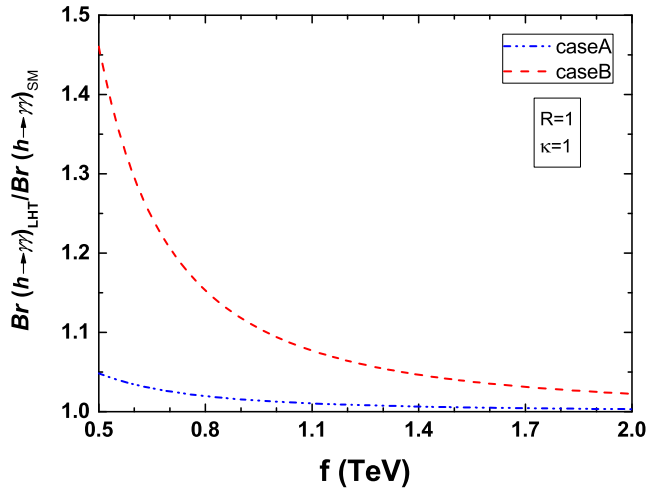}}\vspace{-0.5cm}\caption{ The
ratios $Br(h\rightarrow \gamma\gamma)_{\textrm{LHT}}/Br(h\rightarrow
\gamma\gamma)_{\textrm{SM}}$ versus the scale $f$ for two
cases.}\label{br}
\end{figure}

In order to investigate the observability, we will perform the Monte
Carlo simulation and explore the sensitivity of $pp \to thj$ at 14
TeV LHC through the channel
\begin{equation}
pp\rightarrow t(\rightarrow \ell^{+}\nu b)h(\rightarrow
\gamma\gamma)j
\end{equation}
where $j$ denotes the light jets and $\ell=e,\mu$. At tree level,
the signal has three particles (a top quark, a Higgs boson and a
forward quark jet) in the final state, which is characterized by
appearing as a narrow resonance centered around the Higgs mass.
Since the contribution from Fig.1(b) is negligible, the difference
for the case A and case B in this signal will mainly come from the
branching ratio of $h\rightarrow \gamma\gamma$. We employ the
effective Higgs-photon-photon coupling \cite{efhaa} and calculate
the branching ratio of $h\rightarrow \gamma\gamma$ by using the
package \textsf{HiggsSignals-1.4.0}\cite{HiggsSignals}. We show the
ratios $\textrm{Br}(h\rightarrow
\gamma\gamma)_{\textrm{LHT}}/\textrm{Br}(h\rightarrow
\gamma\gamma)_{\textrm{SM}}$ as a function of the scale $f$ for two
cases in Fig.\ref{br}. We can see that the Br($h\rightarrow
\gamma\gamma$) for case B is larger than that for case A, this is
because the $hb\bar{b}$ coupling in the LHT-B is suppressed much
sizably so that the branching ratio of $h\rightarrow \gamma\gamma$
is enhanced greatly. Besides, the Br($h\rightarrow \gamma\gamma$) in
the LHT model is larger than that in the SM and tends to the SM
value with the scale $f$ increasing, which means the LHT effect
decouples as the scale $f$ increasing.

The backgrounds can be divided into two classes according to their
resonant or nonresonant behavior in the diphoton system:

(i) resonant backgrounds, including $Whjj$, $Zhjj$ and $t\bar{t}h$,
these processes have a Higgs boson decaying to two photons in the
final states;

(ii) nonresonant backgrounds, including $t\bar{t}\gamma\gamma$,
$tj\gamma\gamma$ and $Wjj\gamma\gamma$, where the $Wjj\gamma\gamma$
production can mimic the signal when one light jet is mistagged as a
$b$ jet.

In our analysis, the backgrounds $Whjj$ and $Zhjj$ are ignored due
to their small production cross sections. We generate the signal and
background events at the parton level by
$\textsf{MadGraph5}$\cite{mad5}, where the $t\bar{t}h$ and
$t\bar{t}\gamma\gamma$ events are selected in the single-lepton
channel of $t\bar{t}$ decay. The \textsf{NNPDF23LO1}\cite{cteq}
parton distribution functions are chosen for our calculations. We
set the renormalization scale $\mu_{R}$ and factorization scale
$\mu_{F}$ of the signal process to be $\mu_{R} = \mu_{F} = (m_{t} +
m_{h})/2$, which can be set analogously in the background processes.
The relevant SM input parameters are taken as follows \cite{pdg}:
\begin{align}
m_t = 173.07{\rm ~GeV},\quad &m_{Z} =91.1876 {\rm ~GeV}, \quad m_h
=125 {\rm ~GeV}, \\
\nonumber \sin^{2}\theta_W = 0.231,\quad &\alpha(m_Z) = 1/128, \quad
\alpha_{s}(m_Z)=0.1184.
\end{align}

We feed the events into \textsf{Pythia} \cite{PYTHIA} for parton
showering and hadronization. Then, we perform a fast detector
simulations by \textsf{Delphes} \cite{Delphes}, where the
(mis)tagging efficiencies are taken as the default values. The
subsequent simulations are performed by $\textsf{MadAnalysis
5}$\cite{madanalysis}. We chose the basic cuts as follows:
\begin{eqnarray}\label{basic}
\nonumber\Delta R_{ij} &>&  0.4\ ,\quad  i,j = \gamma, \ell, b~\text{or}\ j  \\
 p_{T}^\gamma &>& 10 \ \text{GeV},\quad |\eta_\gamma| <2.5 \\
\nonumber  p_{T}^\ell &>& 10 \ \text{GeV}, \quad  |\eta_\ell|<2.5  \\
\nonumber  p_{T}^{b} &>& 20 \ \text{GeV}, \quad  |\eta_{b}|<2.5  \\
\nonumber  p_{T}^j &>& 20 \ \text{GeV},\quad  |\eta_j|<5.
\end{eqnarray}

\begin{figure}[htbp]
\scalebox{0.4}{\epsfig{file=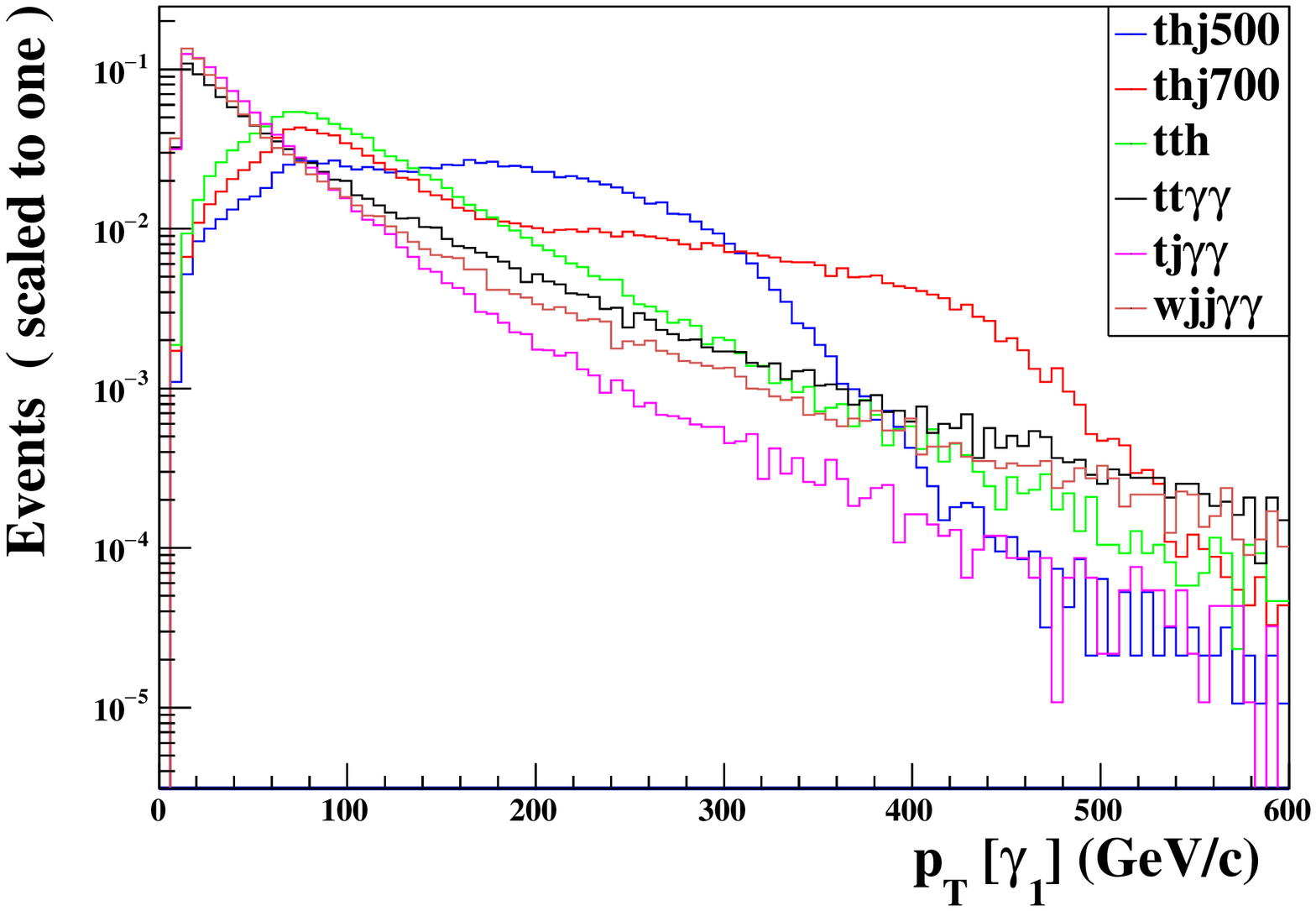}}\hspace{-0.cm}
\scalebox{0.4}{\epsfig{file=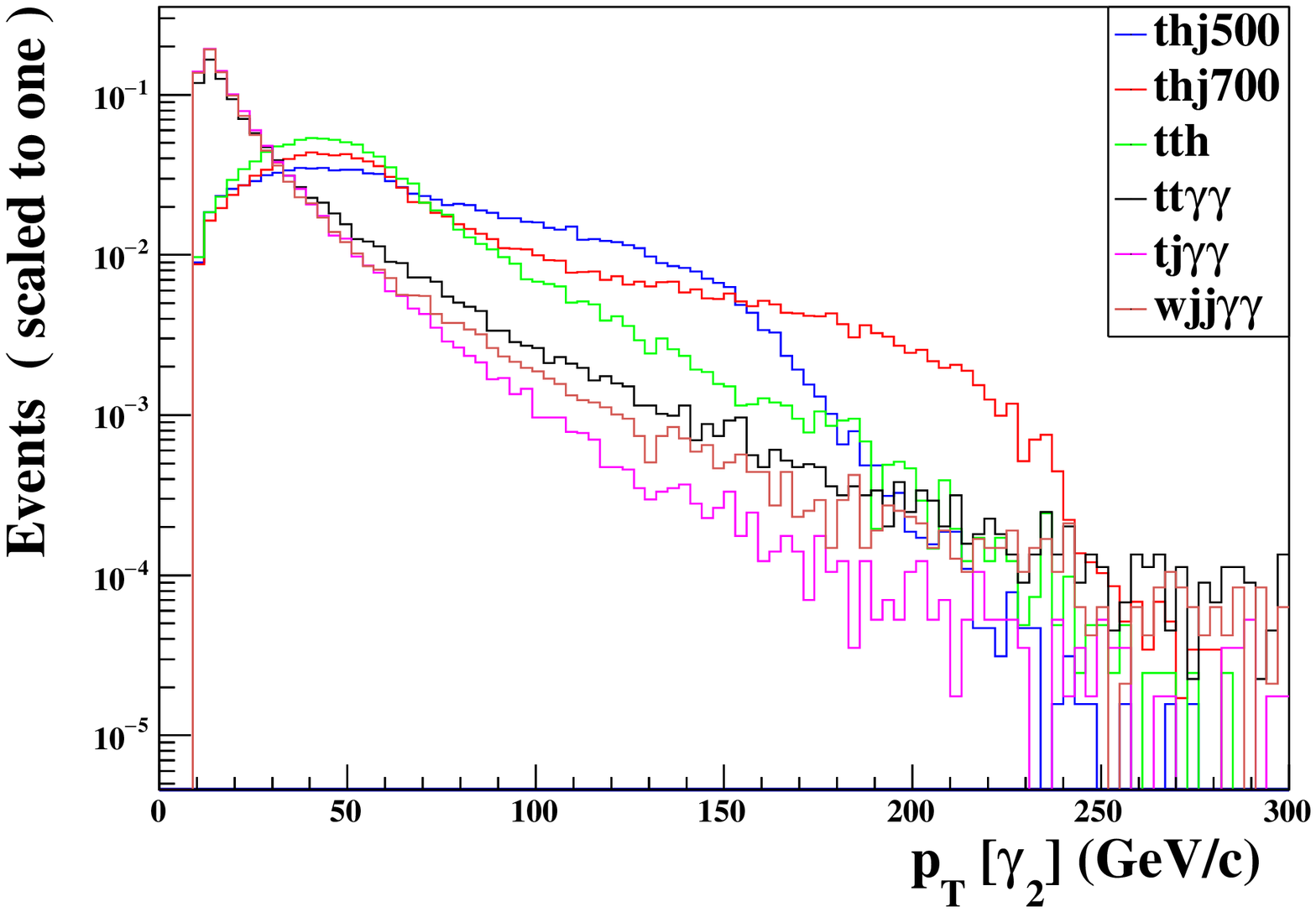}}\vspace{-0cm}
\scalebox{0.4}{\epsfig{file=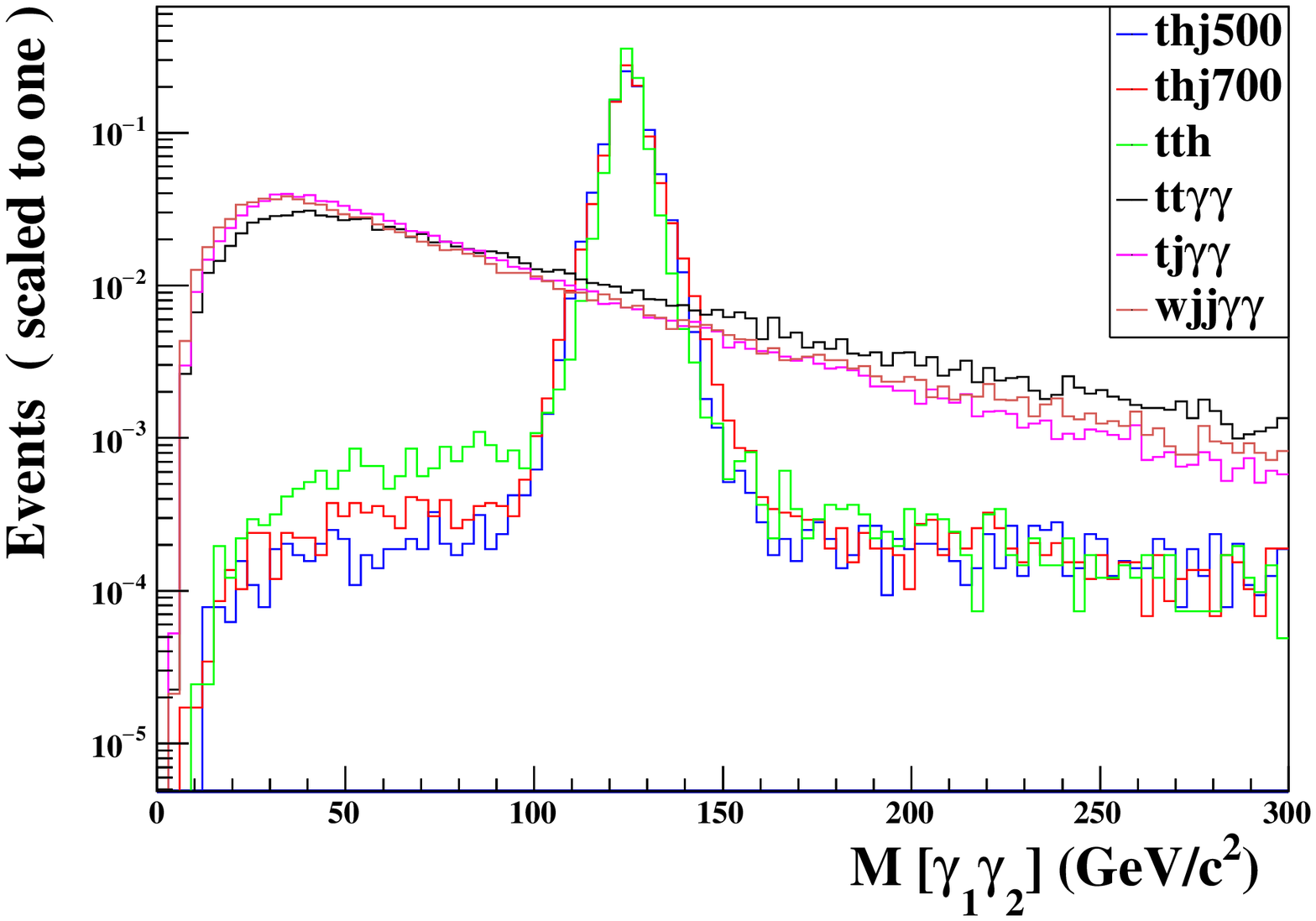}}\hspace{-0.cm}
\scalebox{0.4}{\epsfig{file=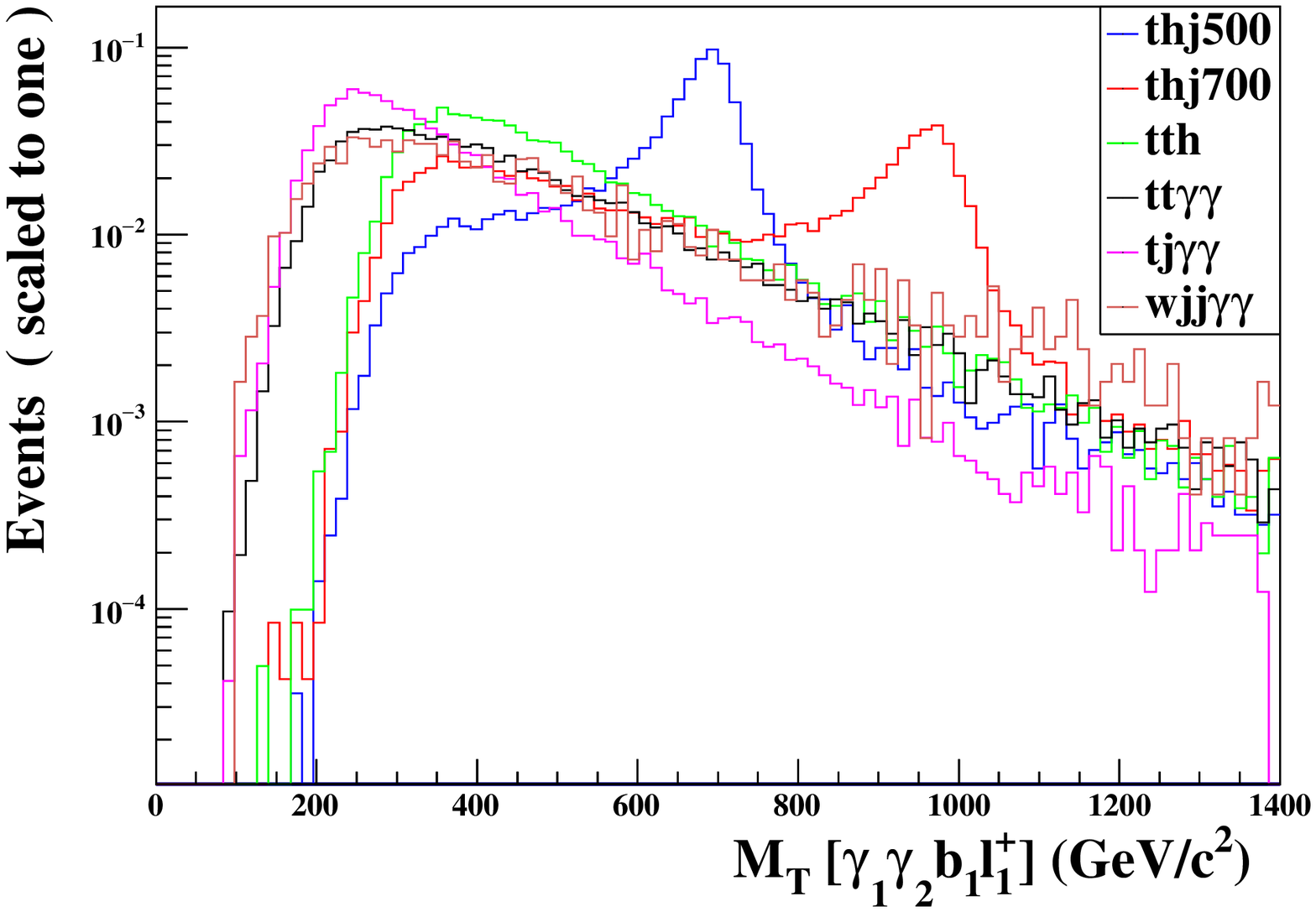}}\vspace{-0.5cm} \caption{The
normalised distributions of $p_{T}^{\gamma_{1}}, p_{T}^{\gamma_{2}},
M_{\gamma_{1}\gamma_{2}}, M_{\gamma_{1}\gamma_{2}b_{1}l_{1}^{+}}$
after the basic cuts in the signal and backgrounds at 14 TeV LHC for
$R=1$.}\label{mm}
\end{figure}

In order to reduce the backgrounds and enhance the signal, some
additional cuts of kinematic distributions are needed. In
Fig.\ref{mm}, we display the normalised transverse momentum
distributions $p_{T}^{\gamma_{1}}, p_{T}^{\gamma_{2}}$ of two
photons, the normalised invariant mass distribution
$M_{\gamma_{1}\gamma_{2}}$ of two photons, the normalised transverse
mass distribution $M_{T}(\gamma_{1}\gamma_{2}b_{1}l_{1}^{+})$ of the
$\gamma_{1}\gamma_{2}b_{1}l_{1}^{+}\met$ system in the signal and
backgrounds at 14 TeV LHC for $R=1$, where the transverse mass
$M_{T}$ is defined as
\begin{eqnarray}
M_T^{2}\equiv\left(\sqrt{(p_{\ell}+p_{b})^{2}+|\vec{p}_{T,\ell}+\vec{p}_{T,b}|^{2}}+|\slashed
p_T| \right)^{2}-|\vec{p}_{T,\ell}+\vec{p}_{T,b}+\vec{\slashed
p}_T|^{2},
\end{eqnarray}
where $\vec{p}_{T,\ell}$ and $\vec{p}_{T,b}$ are respectively the
transverse momentums of the charged leptons and $b$-quark, and
$\slashed p_T$ is the missing transverse momentum derived from the
negative sum of visible momenta in the transverse direction.

Since the two photons in the signal and the resonant backgrounds
come from the Higgs boson, they have the harder $p_{T}$ spectrum
than those in the non-resonant backgrounds. Thus, we can apply the
cuts of two photons to suppress the non-resonant backgrounds. Next,
the signal and the resonant backgrounds have the diphoton
invariant-mass peak at $m_{h}$, which can be used to further reduce
the non-resonant backgrounds. Besides, due to the resonance effect
of the top partner $T_{+}$, the transverse mass distribution
$M_{T}(\gamma_{1}\gamma_{2}b_{1}l_{1}^{+})$ have endpoints round
$m_{T_{+}}$ in the signal, which can be used to remove the
backgrounds effectively.

According to the above analysis, we require the events after the
basic cuts to satisfy the following criteria:
\begin{eqnarray}
&&\textrm{Cut-1}: p_{T}^{\gamma_{1}}>60\textrm{GeV}, p_{T}^{\gamma_{2}}>30\textrm{GeV};\nonumber\\
&&\textrm{Cut-2}: |M_{\gamma\gamma}-m_h|<10\textrm{GeV};\nonumber\\
&&\textrm{Cut-3}:
M_{\gamma_{1}\gamma_{2}b_{1}l_{1}^{+}}>550\textrm{GeV}.\nonumber
\end{eqnarray}

In Table.\ref{cutflow}, we summarize the cut-flow cross sections of
the signal and backgrounds after imposing the cuts. For comparison,
we chose two sets of benchmark points that ($f=500$ GeV, $R=1$)
correspond to $m_{T_{+}}=702$ GeV and ($f=700$ GeV, $R=1$)
correspond to $m_{T_{+}}=984$ GeV, which satisfy the constraint of
the current Higgs data and the electroweak precision
observables(EWPO)\cite{constraint}. As we know, the process $pp \to thj$
exists in the SM. The leading-order cross section for this process
is $\sigma_{\textrm{SM}}^{14\textrm{TeV}}(pp \to thj)$ = 80.4
fb\cite{thjsm}, which retains about 0.02 fb after the subsequent
decays and the basic cuts.

\begin{table}[ht!]
\fontsize{12pt}{8pt}\selectfont \caption{Cutflow of the cross
sections for the signal and backgrounds at 14 TeV LHC on the
benchmark points ($f=500$ GeV, $R=1$) and ($f=700$ GeV, $R=1)$ for
two cases. All the conjugate processes of the signal and backgrounds
have been included.\label{cutflow}}
\begin{center}
\newcolumntype{C}[1]{>{\centering\let\newline\\\arraybackslash\hspace{0pt}}m{#1}}
{\renewcommand{\arraystretch}{1.5}
\begin{tabular}{ |C{0.3cm} C{0.3cm} |C{1.5cm} |C{1.5cm}|C{1.5cm} |C{1.5cm}|C{1cm}C{1.2cm} C{1.2cm} C{1.5cm}|}
\cline{1-8} \hline
&\multicolumn{1}{c|}{\multirow{3}{*}{Cuts}}&\multicolumn{8}{c|}{$\sigma$($\times
10^{-3}$fb)}\\\cline{3-10}
&&\multicolumn{2}{c|}{Signal-caseA}&\multicolumn{2}{c|}{Signal-caseB}
&\multicolumn{4}{c|}{Backgrounds}
\\\cline{3-10}
&&$thj500$&$thj700$&$thj500$&$thj700$
&$t\overline{t}h$&$t\overline{t}\gamma\gamma$&$tj\gamma\gamma$&$Wjj\gamma\gamma$\\\cline{1-10}
\multicolumn{2}{|c|}{\multirow{1}{*}{Basic cuts}}&65.7
&33.7&91.5&39.8&204&4462&3091&35770
\\\hline
\multicolumn{2}{|c|}{\multirow{1}{*}{$ p_{T}^{\gamma_{1}}>60\text{GeV}$}}&\multirow{2}{*}{34.6}&\multirow{2}{*}{15.9}&\multirow{2}{*}{48.2}&\multirow{2}{*}{18.8}&\multirow{2}{*}{71.5}&\multirow{2}{*}{665.8}&\multirow{2}{*}{411.1}&\multirow{2}{*}{4494}\\
\multicolumn{2}{|c|}{\multirow{1}{*}{$
p_{T}^{\gamma_{2}}>30\text{GeV}$}}&&&&&&&&\\\hline
\multicolumn{2}{|c| }{\multirow{1}{*}{ $
|M_{\gamma\gamma}-m_h|<10$GeV}}&29.8
&13.7&41.5&16.2&66.1&65.9&45.2&449.4\\\hline\multicolumn{2}{|c|}{\multirow{1}{*}{
$M_{\gamma_{1}\gamma_{2}b_{1}l_{1}^{+}}>$550GeV }}
&10.3&3.3&14.3&3.9&9.9& 8.6&4.0&8.5
\\
\hline
\end{tabular}}
\end{center}
\end{table}

\begin{figure}[htbp]
\begin{center}
\scalebox{0.35}{\epsfig{file=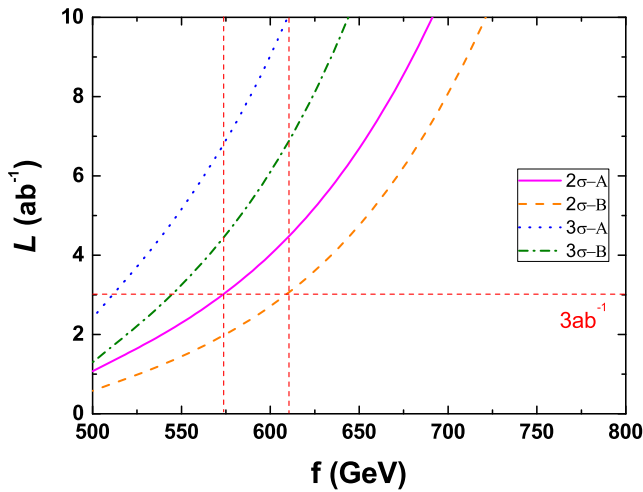}}\vspace{-0.5cm}\hspace{-0.cm}
\caption{The statistical significance of $pp\rightarrow
t(\rightarrow \ell^{+}\nu b)h(\rightarrow \gamma\gamma)j$ at 14 TeV
LHC on the $L\sim f$ plane for $R=1$ in two cases. The contribution
of the charge conjugate process has been included.}\label{ssl}
\end{center}
\end{figure}
To estimate the observability quantitatively, the Statistical
Significance ($SS$) is calculated after final cut by using Poisson
formula\cite{Poisson}
\begin{eqnarray}
SS=\sqrt{2L\left [ (S+B)\ln\left(1+\frac{S}{B}\right )-S\right ]},
\end{eqnarray}
where $S$ and $B$ are the signal and background cross sections and
$L$ is the integrated luminosity. The results of the $SS$ values
depending on the integrated luminosity $L$ at 14TeV LHC for $R=1$ in two cases are shown in Fig.\ref{ssl}, where the contours of
$2\sigma$ and $3\sigma$ significance are displayed. We can see that
the scale $f$ can be excluded up to 573 GeV (correspond to
$m_{T_{+}}$=805 GeV) for case A and 610 GeV (correspond to
$m_{T_{+}}$=857 GeV) for case B at $2\sigma$ confidence level at 14
TeV LHC with the integrated luminosity $L = 3 \rm ab^{-1}$. If the
integrated luminosity can reach $10 \rm ab^{-1}$, the $2\sigma$
exclusion limit of the scale $f$ will be pushed up to 690 GeV
(correspond to $m_{T_{+}}$=970 GeV) for case A and 720 GeV
(correspond to $m_{T_{+}}$=1012 GeV) for case B.

For the process $pp\rightarrow
thj$, comparing the result of decay $h\to \gamma\gamma$ with that of decay $h\to b\bar{b}$ in Ref.\cite{thjLHT}, we can see that the limits on the $T_{+}$ mass from $h\to \gamma\gamma$ are weaker than that from $h\to b\bar{b}$.  For the result of decay $h\to b\bar{b}$, it is worth noting that the limit on $m_{T_{+}}$ for case B is lower than that for case A, which is because the smaller bottom Yukawa coupling in case B (cf. Eq.(9)) leads to a higher suppression of the branching ratio of $h\to b\bar{b}$. By contrast, the branching ratio of $h\rightarrow \gamma\gamma$ for case B is larger than that for case A as shown in Fig.\ref{br}.
As a result, the limits on $m_{T_{+}}$ from the $thj$ production with $h\rightarrow \gamma\gamma$ for the two cases are just the reverse. 

Besides, the limits on the $T_{+}$ mass from the global fit of the Higgs data and EWPO have been performed in Ref.\cite{constraint}, where the $T_{+}$ mass can be excluded up to 920(750) GeV for case A(B). Combining the case A and case B, we can see that the limits on the $T_{+}$ mass from the $thj$ production with $h\rightarrow \gamma\gamma$ at LHC with High-Luminosity can be comparable with the global fit of the Higgs data and EWPO.

Recently, the direct searches for the vector-like $T$ at 13 TeV
LHC have been performed by ATLAS\cite{ATLAS-T} and CMS\cite{CMS-T} Collaborations relying on signatures induced by both the vector-like $T$ pair-production and single-production modes, corresponding to an integrated luminosity of about 36 fb$^{-1}$. For the pure $Wb$ decay, the observed (expected) 95\% CL lower limits on the $T$ mass are 1350 GeV (1310GeV). For the pure $Zt$ decay, the observed (expected) 95\% CL lower limits on the $T$ mass are 1160 GeV (1170 GeV). However, their bounds strongly depend on the assumptions on the decay branching ratios and the properties of the top partner, especially its group representations. In the LHT model, these limits can be relaxed due to the impure decay modes of the top partner $T_{+}$ and the specific group representations, and the detailed confirmation will
require the Monte Carlo simulations of the signals and backgrounds and the comprehensive collider analysis.

\section{Summary}

In the framework of the LHT model, we investigate the observability
of $pp\rightarrow thj$ with decays $t\rightarrow \ell^{+}\nu b$ and
$h\rightarrow \gamma\gamma$ at 14 TeV LHC. By performing a fast
detector simulation, we find that the $thj$ search in the selected
channel can excluded the $m_{T_{+}}$ up to 805 GeV for case A and
857 GeV for case B at $2\sigma$ confidence level at 14 TeV LHC with
the integrated luminosity $L= 3 \rm ab^{-1}$. This excluded region will be further expanded if the higher integrated luminosity is obtained at the LHC.

\section*{Acknowledgement}
This work was supported by the National Natural Science Foundation
of China (NNSFC) under grants No.11405047 and the Startup
Foundation for Doctors of Henan Normal University under Grant
No.qd15207.

\end{document}